\documentclass[a4paper,twocolumn, superscriptaddress,prd]{revtex4-1}

\usepackage[utf8]{inputenc}
\usepackage[english]{babel} 
\usepackage{graphicx}
\usepackage{amssymb,amsfonts,amsmath,braket,physics}
\usepackage{nicefrac}
\usepackage{multirow}
\usepackage{hyperref} 
\usepackage{xcolor}
\usepackage{color}
\definecolor{lightblue}{cmyk}{1,0.3,0,0.2}
\definecolor{purplePink}{cmyk}{0,1,0,0.2}
\hypersetup{colorlinks, bookmarksnumbered, citecolor=lightblue, linkcolor=lightblue, pdfstartview=FitH, urlcolor=purplePink, linktocpage}

\usepackage{array}
\newcommand{\PreserveBackslash}[1]{\let\temp=\\#1\let\\=\temp}
\newcolumntype{C}[1]{>{\PreserveBackslash\centering}p{#1}}
\newcolumntype{R}[1]{>{\PreserveBackslash\raggedleft}p{#1}}
\newcolumntype{L}[1]{>{\PreserveBackslash\raggedright}p{#1}}

\usepackage[nolist]{acronym}

\newcommand{\leri}[1]{\left(#1 \right)}
\usepackage{bbm} 

\begin{document}
\title{Current bounds on baryogenesis from complex Yukawa couplings of light fermions}

\author{Shahaf Aharony Shapira}
\email{shahaf.aharony@weizmann.ac.il}
\affiliation{Department of Particle Physics and Astrophysics,\\
	Weizmann Institute of Science, Rehovot, Israel 7610001}

%---------------------------
\begin{abstract}
%---------------------------
We calculate the contribution to the \ac{BAU} from a $\mathcal{CP}$-violating source of the light quarks (charm, strange, down, up) and the electron, resulting from a dimension-six effective field theory term. We derive relevant bounds from the electric dipole moments of the electron and neutron to estimate the maximal contribution from each single flavor modification. Current bounds show that the charm quark can generate at most $\mathcal{O}(1\%)$ of the \ac{BAU}, while the lighter quarks and the electron contribute at much lower levels.
\end{abstract}

\maketitle	
\acresetall

%---------------------------
\begin{acronym}
	\acro{BAU}{baryon asymmetry of the Universe}
	\acro{SM}{Standard Model}
	\acro{NP}{New Physics}
	\acro{EDM}{Electric Dipole Moment}
	\acro{SMEFT}{Standard Model effective field theory}
	\acro{CPV}{$\mathcal{CP}$-violating}
	\acro{VEV}{vacuum expectation value}
	\acro{LHC}{Large Hadron Collider}
	\acro{eEDM}{electron-Electric Dipole Moment}
	\acro{nEDM}{Neutron-Electric Dipole Moment}
	\acro{dim-$6$}{dimension-six}
	\acro{dim-$4$}{dimension-four}
	\acro{C.L.}{confidence level}
	\acro{U.B.}{upper bound}
	\acro{ggF}{gluon-gluon fusion}
\end{acronym}

%---------------------------
\paragraph*{Introduction}
%---------------------------
The \ac{BAU} is defined and measured~\cite{Planck:2015fie,Tanabashi:2018oca} to be
\begin{align}
Y_B &\equiv \frac{n_{B} - n_{\bar{B}}}{s} \approx \leri{8.6 \pm 0.1} \cdot 10^{-11} \equiv Y_B^\text{obs}\,,
\end{align} 
where $n_{(\bar{B}) B}$ is the (anti-)baryon number density and $s$ is the entropy density of the Universe. A non-vanishing value can be either the result of initial conditions, or dynamically generated during the early Universe. The former requires fine tuning and is inconsistent with inflation. The latter, which is the more acceptable mechanism to address the asymmetry, is called Baryogenesis (See~\cite{Cline:2006ts,Morrissey:2012db}, for reviews).

There are three necessary conditions, known as the Sakharov conditions~\cite{Sakharov:1967dj}, that are required from any theory in order to explain such an imbalance: Baryon number violation, $\mathcal{C}$-symmetry and $\mathcal{CP}$-symmetry violation and interactions out of thermodynamic equilibrium. Although the \ac{SM} meets all three criteria, the rate at which it contributes is far too small to account for the observed baryon asymmetry~\cite{Gavela:1993ts,Huet:1994jb} due to two factors: the smooth crossover of the electroweak phase transition and  the suppression from the Kobayashi-Maskawa (KM) mechanism of $\mathcal{CP}$ violation.
Thus, if the baryon asymmetry was generated via electroweak baryogenesis, the electroweak phase transition had to be strongly first order and a new sources of $\mathcal{CP}$ violation must exist at, or at least not far above, the electroweak scale.

\ac{NP} beyond the \ac{SM} is highly motivated by several open questions in Physics (e.g. dark matter, neutrino masses). However, despite the efforts made to discover new particles, none were found up to the TeV scale~\cite{CMS:2019ybf,ATLAS:2019fgd}. It is then plausible that \ac{NP} is above the electroweak scale, and thus could be integrated out. This allows us to use \ac{SMEFT} tools to explore higher order terms, without being model-dependent.\\

We add a \ac{CPV} phase using a \ac{dim-$6$} coupling of three Higgs fields to the \ac{SM} charged fermions. The \ac{BAU} is then proportional to the \ac{CPV} source, which could be constrained by the \ac{EDM} of both the electron and neutron and by the Higgs boson decay and production rates. This was previously done for the third generation particles~\cite{deVries:2017ncy,deVries:2018tgs,Fuchs:2020uoc} and the muon~\cite{Fuchs:2019ore}. Of that list, it was shown that the $\tau$ is the only sole-contributor that can provide the entire observed value of the \ac{BAU}~\cite{deVries:2018tgs,Fuchs:2020uoc}. We applied this procedure to evaluate the contribution from all of the \ac{SM} particles, including the light quarks (charm, strange, down, up) and the electron, and discuss the results here.\\

This Letter is organized as follows. First, we describe the \ac{SMEFT} framework, including the complex \ac{dim-$6$} term and the \ac{CPV} source it generates. We then outline key points in the process of electroweak baryogenesis, which are formulated by the two-step approach via the transport equations followed by the sphaleron process. Next, we present our numerical results for the contribution of a single flavor to the \ac{BAU}. The contribution is later bounded using the experimental measurements of the electron and neutron \ac{EDM}s and of various Higgs boson processes. Finally, we discuss our results and conclusions.\\

%---------------------------
\paragraph*{\ac{SMEFT} framework}
%---------------------------
We examine the implications of adding the following effective \ac{dim-$6$} terms to the \ac{SM},
\begin{align} 
\label{dim6Term}
\mathcal{L}_\text{eff Yuk} &= - \leri{ y_{f} + \frac{|H|^2}{\Lambda^2} \leri{X^R_{f} + i X^I_{f}} } \overline{\psi_{Lf}} \psi_{Rf} H + \text{h.c.} \,, 
\end{align}
where $y_f$ is the \ac{dim-$4$} Yukawa coupling, $H$ is the \ac{SM} Higgs field $H \sim \leri{1,\ 2}_{+1/2} $, $\Lambda$ is the \ac{NP} scale, $X$ is the \ac{dim-$6$} Wilson coefficient and $\psi$ is a \ac{SM} fermion. In our notation, the lower index, $f$, denotes the flavor whereas the upper index distinguishes between the real ($R$) and imaginary ($I$) coefficients. We find it useful to define~\cite{Fuchs:2020uoc,Fuchs:2019ore}
\begin{align}
T^{R,I}_{f} &\equiv \frac{v^2}{2 \Lambda^2} \frac{X^{R,I}_{f}}{y_{f}} \,,
\end{align}
where $v = 246$ GeV is the \ac{VEV} of the Higgs background field ($h$) defined $H = \frac{1}{\sqrt{2}} \begin{pmatrix}  0, & v+h  \end{pmatrix}^T$. Accordingly, the mass ($m$) and effective Yukawa coupling ($\lambda$) of each flavor can be defined and explicitly written as
\begin{align}
\mathcal{L}_\text{eff Yuk}  & \supset - m_{f}\bar{\psi}_{Lf} \psi_{Rf} - \lambda_{f}  \bar{\psi}_{Lf} \psi_{Rf} h + \text{h.c.}  \,,
\end{align}
\begin{align} \label{lambdaDef}
m_{f} &= \frac{v y_{f}}{\sqrt{2}} \leri{1 + T^{R}_{f} + i T^{I}_{f}}\,, &
\lambda_{f} &=\frac{y_{f}}{\sqrt{2}} \leri{1 + 3T^{R}_{f} + i3 T^{I}_{f}} \,,
\end{align}
In the mass basis, where $m_f \in \mathbb{R}$, they are given by~\cite{Fuchs:2020uoc,Fuchs:2019ore}
\begin{align}
m_{f} ' & = \frac{v y_{f}}{\sqrt{2}} \sqrt{\leri{1 + T^{R}_{f}}^2 + {T^{I}_{f}}^2} \,, \label{eqn:defmPrime} \\
\lambda_{f}' & = \frac{y_{f}}{\sqrt{2}} \frac{ \leri{T^{R}_{f} +1} \leri{3 T^{R}_{f} +1} + T^{I}_{f} \leri{3 T^{I}_{f}+2i} }{\sqrt{ \leri{1 + T^{R}_{f}}^2 +{T^{I}_{f}}^2 }}\,. \label{eqn:defLambdaPrime}
\end{align}
Throughout this work we will use the complex parameter $\kappa_f$, representing the deviation from the \ac{SM} in the mass basis,
\begin{align} \label{kappaDef}
\kappa_f &\equiv \frac{\lambda_f' v}{m_f'} = 3 - \frac{2}{1+T_f^R + i T_f^I} \,, \nonumber \\
\kappa_f^I &\equiv \Im(\kappa_f) =  \frac{2 T^I_f}{\leri{1+ T^R_f}^2 + {T^I_f}^2} \,.
\end{align} 
Equivalently, one could express $\kappa_f = \frac{\lambda_f'}{\lambda_f^\text{' SM}}$ where, as in the \ac{SM},  $\lambda_f^\text{' SM} \equiv \frac{m_f'}{v}$. It is convenient to use $\kappa_f $ since the baryon asymmetry is proportional to the \ac{CPV} source ($S_f$, Eqn.~\eqref{sourceTerm}) and therefore linear in  $\kappa_f^I$~\cite{Fuchs:2020uoc},
\begin{align}
Y_B \propto S_f \propto \Im (m_f^* m_f') \propto \kappa_f^I \,.
\end{align} 
In the above, we use the \ac{VEV} insertion approximation~\cite{Lee:2004we}, to leading order.
% As we show below, $\kappa_f^I$ can be constrained by \ac{CPV} observables, such as the \ac{EDM}s, and by Higgs-related measurements in colliders.
\\

\begin{figure}[h]
	\begin{center}
		\includegraphics[]{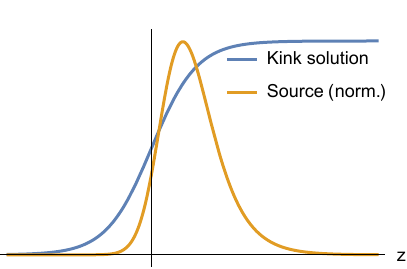}
	\end{center}
	\caption{\textit{Schematic description of the kink solution and its resulting \ac{CPV} source along the distance from the bubble wall ($z=0$). Blue: the kink solution for the background Higgs boson field, as given in Eqn.~\eqref{kink sol}, orange: the resulting source, normalized.} }  \label{plot:schematicSource}
\end{figure}

%---------------------------
\paragraph*{Electroweak baryogenesis}
%---------------------------
By adding a complex effective Yukawa coupling to one of the fermions of the \ac{SM} we introduce a \ac{CPV} source to the model. The source depends on the background Higgs boson field $h$, which acquires a \ac{VEV}. Here we assume that the electroweak phase transition is strongly first order, and describe $h$ by the kink solution (Eqn.~\eqref{kink sol}). This choice results with a \ac{CPV} source that peaks mostly inside the non-vanishing \ac{VEV} bubble. A schematic plot is given in Fig.~\ref{plot:schematicSource}. Namely, the \ac{CPV} source generates a chiral asymmetry, mainly inside the broken phase. 
That asymmetry can be transformed into an abundance of baryons via the weak sphaleron, which is a non-perturbative effect of the \ac{SM}. The rate of the weak sphaleron process is given by $\Gamma_{ws} \sim e^{-\langle h \rangle / T} T$~\cite{Kuzmin:1985mm}, where $T$ is the temperature. Although the rate is exponentially suppressed inside the bubble, it is fast outside the bubble, during the early Universe. Therefore, for baryogenesis to occur at the electroweak phase transition, the chiral asymmetry should have propagated outside the bubble and into the symmetric phase~\cite{Joyce:1994zn}. Finally, as the bubble continues to expand, it eventually captures the resulting baryon asymmetry.\\

%---------------------------
\paragraph*{Two-step approach}
%---------------------------
The \ac{dim-$6$} term described in Eqn.~\eqref{dim6Term} will affect the dynamics of the number densities, which are defined as the difference between the number densities of particles and anti-particles. This effect is the first step of the approach and is described via a set of transport equations. We generalize the set given in~\cite{deVries:2018tgs} to include all the fermions of the \ac{SM} in addition to the Higgs boson,
\begin{align} \label{transEqnFullSet}
\partial_\mu  U_i^\mu \equiv \partial U_i &= - \Gamma_M^{U_i} \mu_M^{U_i} - \Gamma_Y^{U_i} \mu_Y^{U_i} + \Gamma_{ss} \mu_{ss} + S_{U_i}\,, \nonumber \\
\partial D_i &= - \Gamma_M^{D_i} \mu_M^{D_i} - \Gamma_Y^{D_i} \mu_Y^{D_i} + \Gamma_{ss} \mu_{ss} + S_{D_i}\,, \nonumber \\
\partial Q_i &= -\partial U_i  -\partial D_i  \,, \nonumber \\
\partial E_i &= - \Gamma_M^{E_i} \mu_M^{E_i} - \Gamma_Y^{E_i} \mu_Y^{E_i} + S_{E_i} \,, \nonumber \\
\partial L_i &= -\partial E_i \,, \nonumber \\
\partial h &=  \sum_{i=1}^3 \Gamma_Y^{U_i} \mu_Y^{U_i}  -\sum_{i=1}^3  \Gamma_Y^{D_i} \mu_Y^{D_i}  - \sum_{i=1}^3  \Gamma_Y^{E_i} \mu_Y^{E_i}  \,,
\end{align}
where $U_R, D_R, Q_L$ are the \ac{SM} quark fields, $E_R, L_L$ are the \ac{SM} charged lepton fields (the chirality is implicit hereafter), $\Gamma_{M,Y}$ are the relaxation and Yukawa rates, respectively, and $S_f$ are the \ac{CPV} sources. The chemical potentials are given by
\begin{align}
\mu_Y^{U_i} &= \frac{{U_i}}{k_{U_i}} -  \frac{{Q_i}}{k_{Q_i}} - \frac{h}{k_h}\,, &
\mu_M^{U_i} &= \frac{{U_i}}{k_{U_i}} -  \frac{{Q_i}}{k_{Q_i}}  \,,  \nonumber \\
\mu_Y^{D_i} &= \frac{{D_i}}{k_{D_i}} -  \frac{Q_i}{k_{Q_i}} + \frac{h}{k_h} \,,  &
\mu_M^{D_i} &= \frac{{D_i}}{k_{D_i}} -  \frac{{Q_i}}{k_{Q_i}}  \,, \nonumber \\
\mu_Y^{E_i} &= \frac{{E_i}}{k_{E_i}} -  \frac{{L_i}}{k_{L_i}}  + \frac{h}{k_h}  \,,&
\mu_M^{E_i} &= \frac{{E_i}}{k_{E_i}} -  \frac{{L_i}}{k_{L_i}} \,,  \nonumber  \\
\mu_{ss} &= \sum_{i=1}^3 \leri{ \frac{2Q_i}{k_{Q_i}} - \frac{U_i}{k_{U_i}}- \frac{D_i}{k_{D_i}} } \,,
\end{align}
where $k$ counts the finite temperature degrees of freedom. All of the benchmark parameters were taken from Ref.~\cite{Fuchs:2020pun}, and are presented in Appendix~\ref{benchParameters}.
The solution of the transport equations is obtained by reduction of order~\cite{Fuchs:2020pun}, i.e. $N$- second order differential equations are written as a set of $2N$- first order differential equations, and numerical diagonalization. By the second step of the approach the left-handed particles participate in the sphaleron process which generates a baryon asymmetry (Eqn.~\eqref{eqn:sphaleronProcess}). \\
The two-step procedure is summarized in more detail in Appendix~\ref{SolvingTransEqns}.\\

\begin{figure}[h]
	\begin{center}
		\includegraphics[scale=0.52]{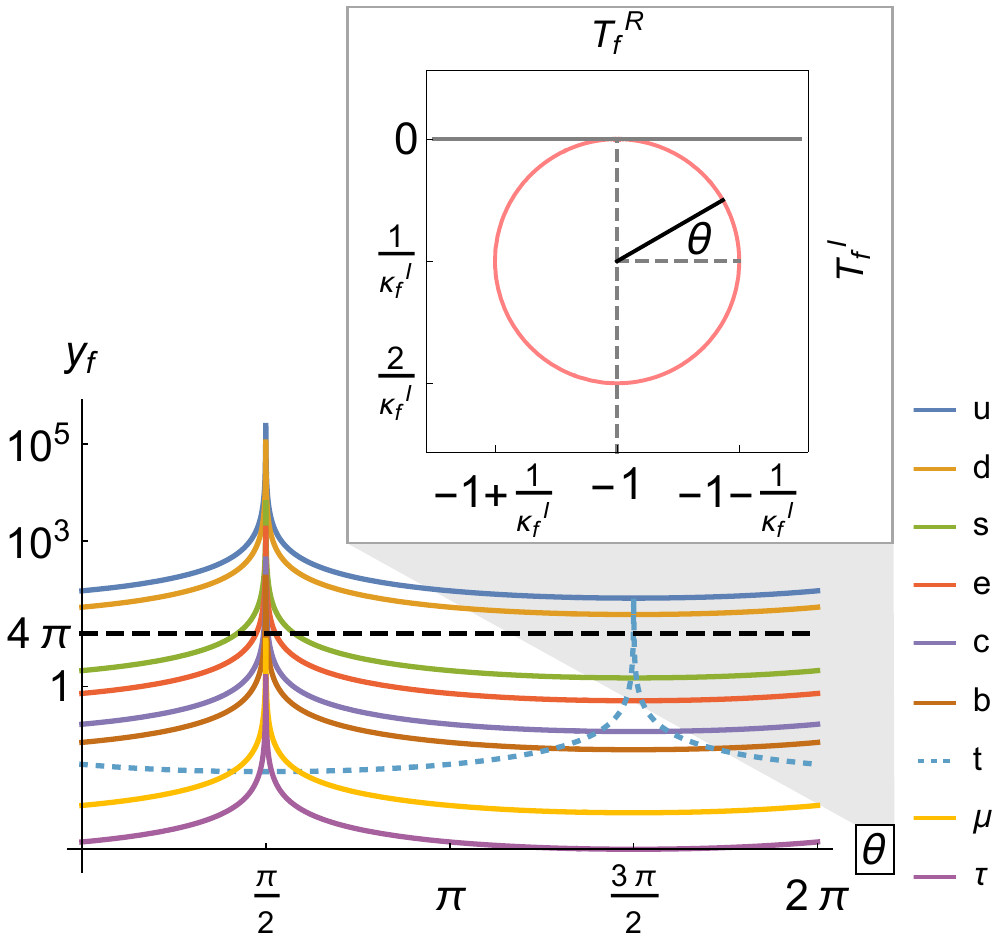}
	\end{center}
	\caption{\textit{\textbf{Main}: Dim-$4$ Yukawa coupling $y_f$ as a function of $\theta$ (See inset). The coupling $y_f$ is expressed by setting the mass $m_f'$ to its measured value and $\kappa_f^I$ to $\kappa_f^{I*}$. The region above the dashed line $y_f = 4 \pi$ is non-perturbative and is therefore considered unfavorable. The up and down quarks are non-perturbative throughout the entire range, which precludes them as a sole-source of the observed \ac{BAU}. \\
			\textbf{Inset}: The geometrical interpretation of Eqn.~\eqref{eqn:circle}, for a given $\kappa_f^I$, is a circle in $\leri{T_f^R, T_f^I}$-space centered around $\leri{-1,\ 1/ \kappa_f^{I} }$ with radius $1/ {\left|\kappa_f^{I} \right|}$. Each point on the circle is mapped to $\theta$, its central angle from the positive horizontal direction. The plot corresponds to $\kappa_f^I < 0$ (See Eqn.~\eqref{eqn:YBcontribution}).} }  \label{plot:ThetaNonPert}
\end{figure}

\paragraph*{Numerical result}
Our numerical calculation for the \ac{BAU} yields
\begin{align} \label{eqn:YBcontribution}
Y_B \leri{\kappa_f^I} = - Y_B^\text{obs}  \cdot & \left(  - 28  \kappa_t^I  + 11  \kappa_\tau^I + 0.2  \kappa_b^I + 0.1 \kappa_\mu^I  \right. \nonumber  \\
& \ \left. +0.03 \kappa_c^I +2 \cdot 10^{-4} \kappa_s^I  +3\cdot 10^{-6}  \kappa_e^I  \right.  \nonumber \\ 
&\  \left. + 4 \cdot 10^{-7} \kappa_d^I +9 \cdot 10^{-8} \kappa_u^I \right) \,. 
\end{align}
This result agrees with previous analysis for the third generation particles~\cite{Fuchs:2020uoc} and the muon~\cite{Fuchs:2019ore}, but also includes the rest of the charged \ac{SM} fermions.\\
We point out that although quarks have (on average) larger \ac{dim-$4$} Yukawa couplings, which positively impact the \ac{CPV} source, they also have more washout. It is the result of lower diffusion~\cite{Cohen:1994ss,Guo:2016ixx} and higher interaction rates, as well as an additional interaction via the strong sphaleron~\cite{Giudice:1993bb}. This important difference makes the charged leptons better candidates for producing the \ac{BAU}, compared to quarks~\cite{Fuchs:2020uoc,deVries:2018tgs,Chung:2009cb}. \\
An interesting feature that holds only for the light fermions is that the ratio between contributions of different flavors of same type (either charged leptons or quarks) to the \ac{BAU} is proportional to the ratio of the mass squared, up to $5 \%$. For $f_1, f_2$ light flavors of same type, we get the following numerical result:
\begin{align}
\frac{Y_B^{f_1}}{Y_B^{f_2}} &\approx \leri{ \frac{m_{f_1}}{m_{f_2}} }^2 \frac{\kappa_{f_1}^I}{\kappa_{f_2}^I}\,.
\end{align}
The mass squared is explicitly introduced to the source term (See Eqn.~\eqref{sourceTerm}), which is otherwise almost identical for same type particles. However, it is non-trivial that the numerical solution of the transport equations is approximately linear with various fermionic sources, given they have different interaction rates. It is a consequence of the negligible difference between the light fermion rates (See Table \ref{tab:interactionRates}) and indeed, this relation does not hold for the third generation particles.\\
Moreover, the solution for every given $\kappa_f^I$ is centered around a cancellation between the \ac{dim-$6$} and \ac{dim-$4$} contributions to $m_f'$. By rearranging the definition of $\kappa_f^I$ (Eqn.~\eqref{kappaDef}), we get
\begin{align} \label{eqn:circle}
\leri{T_f^R + 1}^2 + \leri{T_f^I - \frac{1}{\kappa_f^{I}}}^2 = \leri{\frac{1}{\kappa_f^{I}}}^2\,.
\end{align}
See inset of Fig.~\ref{plot:ThetaNonPert} for the geometrical interpretation. The center of the circle, at $T_f^R = -1$, implies that the mass of the fermion is effectively generated by the imaginary part of the \ac{dim-$6$} term (See Eqn.~\eqref{eqn:defmPrime}). This point requires a fine-tuned cancellation between the \ac{dim-$4$} Yukawa coupling and the real part of the \ac{dim-$6$} term. Furthermore, we do not expect this tension to be relaxed by introducing higher order terms, as they have negligible contribution.\\
Finally, the desired $\kappa_f^{I}$ that saturates the contribution to the \ac{BAU} to its observed value could correspond to an unfavorable solution, when demanding the theory to be perturbative.  Let us denote the single flavor modification $\kappa_f^{I}$ which satisfies $Y_B = Y_B^\text{obs}$, according to Eqn.~\eqref{eqn:YBcontribution}, by $\kappa_f^{I*}$. The solution $\kappa_f^{I*}$ corresponds to a circle in $\leri{T_f^R, T_f^I}$-space. By setting the mass $m_f'$  (See Eqn.~\eqref{eqn:defmPrime}) to its measured value, we calculate the resulting \ac{dim-$4$} coupling $y_f$ for each point on the circle, as a function of its central angle from the positive horizontal direction $(\theta)$. For some cases, depending on the particle and the position on the circle, it requires $y_f > 4 \pi$ which in non-perturbative (See Fig.~\ref{plot:ThetaNonPert}), rendering the analysis moot.\\
A theoretical upper bound on $\left| \kappa_f^I \right| $ is produced by requiring there exists $\theta_p$ for which $y_f \leri{\theta_p} \leq 4 \pi$. Although this constraint is fairly weak, we present the perturbativity bound in Table~\ref{tab:BAUresults} for comparison reasons.\\
Note that $\theta = \frac{\pi}{2} \leri{\frac{3 \pi}{2} \text{ for the top}}$, which corresponds to $\leri{-1,0}$ in $\leri{T_f^R, T_f^I}$-space, is clearly un-physical and should be excluded. For large values of $\kappa_f^{I*}$, the radius of the circle, $1/{\left|\kappa_f^{I*} \right|}$, is too small to escape this critical region. We specifically point out the up and down quarks, for which there is no perturbative theory that can account for the observed \ac{BAU}.\\

%---------------------------
\paragraph*{Bounds}
%---------------------------	
The implications of a non-zero $\kappa_f^I$ are manifold; In addition to the generation of baryon asymmetry, which was discussed above, we focus on the contribution of a single $\kappa_f^I$ to the \ac{EDM}s of the electron and the neutron, and Higgs related measurements~\cite{Chien:2015xha}. In this Section we use these experimental results to constrain the maximal value of $\left|\kappa_f\right|$, assuming $\kappa_{\tilde{f}} = 1$ for all $\tilde{f} \neq f$. This will allow us to infer the maximal contribution, from a single flavor modification, to the \ac{BAU}.
 \\

%---------------------------
\paragraph*{\ac{eEDM} Bound} 
%---------------------------
The \ac{U.B.} obtained by the ACME collaboration on the \ac{eEDM}~\cite{Andreev:2018ayy} is
\begin{align} \label{EDMupperBound}
\left| d_e^\text{max} \right| &= 1.1 \cdot 10^{-29} \ e \ \text{cm at $90\%$~C.L.} \,.
\end{align}

\noindent The contribution of the \ac{SM} fermions, other than the top and the electron, to the \ac{eEDM} is given by~\cite{Brod:2018pli,Panico:2018hal}
	\begin{align} \label{BrodeEDM}
	\frac{d_e}{e} &\simeq 4 N_c Q_f^2 \frac{\alpha}{\leri{ 4 \pi}^3}  \frac{m_e m_f^2}{v^2 m_h^2} \leri{ \ln^2\leri{\frac{m_f^2}{m_h^2}} + \frac{\pi^2}{3} } \kappa_f^I \,.
	\end{align}
	
	\noindent The contribution of the top quark to the \ac{eEDM} is given by~\cite{Brod:2013cka}
	\begin{align} \label{BrodTopeEDM}
	\frac{d_e}{e} &\simeq 9.4  \cdot 10^{-27} \kappa_t^I  \  \text{cm} \,.
	\end{align}
	
	\noindent The contribution of the electron to the \ac{eEDM} is given by~\cite{Altmannshofer:2015qra}
	\begin{align} \label{eEDMAltmannshofer}
	\frac{d_e}{e} & \simeq  5.1  \cdot 10^{-27} \kappa_e^I \  \text{cm} \,.
	\end{align}

%---------------------------
\paragraph*{\ac{nEDM} bound} 
%---------------------------
The \ac{U.B.} obtained by the \ac{nEDM} collaboration on the \ac{nEDM}~\cite{Abel:2020gbr} is
\begin{align} \label{nEDMupperBound}
\left| d_n^\text{max} \right| &= 1.8 \cdot 10^{-26} \ e \ \text{cm at $90\%$~C.L.} \,.
\end{align}
This bound constrains only the $\kappa_f^I$ of the quarks. The implications of the \ac{nEDM} bound on $\kappa^I_{u,d,s}$ were calculated in~\cite{Brod:2018lbf}. Here we show the $90\%$~\ac{C.L.} bound, updated to the latest measurement:
\begin{align}
\left| \kappa_u^I \right| & \lesssim 0.6 \,, &
\left| \kappa_d^I \right| & \lesssim 0.14 \,, &
\left| \kappa_s^I \right| & \lesssim 4.5 \,.
\end{align}
We also update the $68\%$~\ac{C.L.} bounds on $\kappa_{b,c}^I$ given in~\cite{Brod:2018pli}, assuming $\kappa^R_{b,c} = 0$,
\begin{align}
\left| \kappa_b^{I} \right| &\lesssim\ 7.4 \,, &
\left| \kappa_c^I \right| & \lesssim 15.5  \,.
\end{align} 
We used the weakest constraint, given for negative sign of the Weinberg-operator, with the short-distance theory uncertainty (in quadrature) for the charm (bottom). \\

%---------------------------
\paragraph*{Collider constraints}
%---------------------------
The signal strength $\mu_{h \to f \bar{f}}$ can be written in terms of the production rate $\sigma$ and branching ratio $\mathcal{B}$ as
\begin{align}
\mu_{i,h \to f \bar{f}} &\equiv  \frac{\sigma_i \leri{pp \to h} \mathcal{B}\leri{h \to f \bar{f}}}{\sigma_i^\text{SM} \leri{pp \to h} \mathcal{B}^\text{SM} \leri{h \to f \bar{f}}} \,.
\end{align} 
We first use measurements of $\mu_{h \to \ell \bar{\ell}}$ to constraint $|\kappa_{\ell}|$ directly, for the charged leptons and bottom quark, denoted $\ell = \tau, \mu, e, b$. Since the contribution of light fermions to the production rate of the Higgs boson is insignificant already at \ac{dim-$4$}, we can safely neglect their effect via the \ac{dim-$6$} term. Regarding the bottom quark, this approximation neglects its $1\%$ loop contribution to \ac{ggF}~\cite{LHCHiggsCrossSectionWorkingGroup:2016ypw}. However, the \ac{eEDM} bound, in this case, turns out to be more significant~\cite{Fuchs:2020uoc}. We therefore approximate
\begin{align}
	\mu_{h \to \ell \bar{\ell}} &\approx \frac{\mathcal{B}\leri{h \to \ell \bar{\ell}}}{ \mathcal{B}^\text{SM} \leri{h \to \ell \bar{\ell}}} \,.
\end{align}
It is then straightforward to translate the \ac{U.B.} of the signal strength to the maximal value of $\left| \kappa_\ell \right|$ using~\cite{Altmannshofer:2015qra}
\begin{align} \label{eqn:muFormula}
	\mu_{h \to \ell \bar{\ell}} &= \frac{\left|\kappa_\ell \right|^2 }{1+ \leri{\left|\kappa_ \ell \right|^2 -1} \mathcal{B} \leri{h \to  \ell \bar{\ell}}^\text{SM}} \,.
\end{align}
The next class is that of the light quarks $q = u,c,d,s$, which could only be bounded via its effect on the total decay width of the Higgs boson. When \ac{NP} interacts only with $q$, i.e. $\kappa_{f} = 1$ for all $f \neq q$, the signal strength of $f$ is modified as
\begin{align} \label{eqn:muFormulab}
	\mu_{h \to {f} \bar{f}} &=\frac{1 }{1+ \leri{\left|\kappa_ q \right|^2 -1} \mathcal{B} \leri{h \to  q \bar{q}}^\text{SM}} \,.
\end{align}
As the lower bound of the signal strength $\mu_{h \to {f} \bar{f}}$ tends to one, the upper bound on $\left| \kappa_q \right|$ gets stronger. Currently, the experimental lower bound closest to unity is that of the bottom, $\mu_{h \to b \bar{b}} = 1.04 \pm 0.20$~\cite{Sirunyan:2018kst} (see Table~\ref{tab:colliderLimits}). Therefore, we use $\mu_{h \to b \bar{b}}$ to solve the above equation and constrain $\left|\kappa_ q \right|$.\\
Lastly, the single flavor modification of the top quark is bounded using the dominant, top mediated, production modes of the Higgs boson: \ac{ggF} and $t\bar{t}h$. The top affects both the production rate $\leri{\sigma/\sigma^\text{SM} = \left| \kappa_t \right|^2}$, as well as the total decay width of the Higgs boson (Eqn.~\eqref{eqn:muFormulab}). The overall effect can be written as~\cite{Fuchs:2020uoc}
	\begin{align}\label{eqn:muFormulac}
		\mu_{\text{ggF}+t\bar{t}h} &= \frac{ \left| \kappa_t \right|^2 }{1+ \leri{\left|\kappa_ t \right|^2 -1} \mathcal{B} \leri{h \to g \bar{g}}^\text{SM}} \,.
	\end{align}

\onecolumngrid

\begin{table}[b]
	\caption{ \it The \ac{BAU} calculated following the full set of transport equations. $Y_B^f$ is the \ac{BAU} resulting from a single source $S_f$. Collider constraints are at $\sim 95\%$~\ac{C.L.}~\cite{Fuchs:2020uoc,Aad:2020xfq,ICHEP:2020,Sirunyan:2018kst,LHCHiggsCrossSectionWorkingGroup:2016ypw,LHCHiggsCrossSectionWorkingGroup:2011wcg,ATLAS:2019old} (for details see Table~\ref{tab:colliderLimits}). \ac{EDM} constraints are at $90\%$~\ac{C.L.}~\cite{Andreev:2018ayy,Abel:2020gbr} for all, except for the bottom and charm, for which the \ac{nEDM} constraints are at $68\%$~\ac{C.L.}~\cite{Brod:2018pli}. Perturbativity bounds are calculated by setting $m_f'$ to its measured value and demanding that there exists $\theta_p$ for which $y_f (\theta_p) = 4 \pi$.}
	\label{tab:BAUresults}
	\begin{center}
		\renewcommand{\arraystretch}{1.5}
		\begin{tabular}{C{1cm} C{2.5cm} C{2cm} C{2cm} C{1.5cm}  C{2cm}  C{2.3cm}} 
			\hline \hline
			\rule{0pt}{1.0em}
			$S_f,$ & \ac{BAU}, $Y_B^f$ & Collider & \ac{eEDM} & \ac{nEDM}& Perturbativity & ${Y_B^f}^\text{max} $ \\ 
			$f$ & $\times \kappa_f^I $ &  $\left| \kappa_f \right|^\text{max}$ & $\left| \kappa_f^{I} \right|^\text{max} $ & $\left| \kappa_f^{I} \right|^\text{max} $ &  $\left| \kappa_f^{I} \right|^\text{max} $ & $ \times Y_B^\text{obs}$ \\[2pt] \hline\hline
			
			$\tau$ & $-9.9 \cdot 10^{-10} $ & $\mathbin{\color{gray} 1.1} $ & $0.3$ & - & $\mathbin{\color{gray}  2 \cdot 10^3}$ & $3.37$ \\ \hline
			
			$\mu$ & $-1.0 \cdot 10^{-11}$ & $1.3$ & $\mathbin{\color{gray} 31} $& {-} & $\mathbin{\color{gray}  4 \cdot 10^4}$ & $0.16$ \\ \hline
			
			$b$ & $-2.1 \cdot 10^{-11} $ & $ \mathbin{\color{gray} 1.7}$	& $0.2$ & $\mathbin{\color{gray} 7.4 } $ & $\mathbin{\color{gray}  1 \cdot 10^3}$ & $5.8 \cdot 10^{-2} $ \\ \hline
			
			$t$ & $+ 2.4 \cdot 10^{-9}$ &$\mathbin{\color{gray} 1.1}$ &  $1.2 \cdot 10^{-3}$ & 
			{-} & $\mathbin{\color{gray}  25}$  &  $3.3 \cdot 10^{-2}$ \\ \hline
			
			$c$ & $-2.7 \cdot 10^{-12} $ & $ \mathbin{\color{gray} 3.9}$	& $ 0.4$ & $\mathbin{\color{gray} 15.5 } $ & $\mathbin{\color{gray}  3 \cdot 10^3}$ & $1.1 \cdot 10^{-2}$ \\ \hline
			
			$s$ & $-1.6 \cdot 10^{-14} $ & $\mathbin{\color{gray} 30}$ & $\mathbin{\color{gray} 109} $ & $ 4.5$ & $\mathbin{\color{gray}  5 \cdot 10^4}$  & $8 \cdot 10^{-4}$ \\ \hline
			
			$d$ & $-3.8 \cdot 10^{-17} $ &$\mathbin{\color{gray}621}$ & $\mathbin{\color{gray} 2.3 \cdot 10^4} $ & $0.14$ &$\mathbin{\color{gray}  9 \cdot 10^5}$ & $ 6 \cdot 10^{-8} $ \\ \hline
			
			$u$ & $-8.2 \cdot 10^{-18} $ &$\mathbin{\color{gray} 1326}$ & $\mathbin{\color{gray} 2.2 \cdot 10^4} $  &$0.6$ & $\mathbin{\color{gray}  2 \cdot 10^6}$  & $ 6 \cdot 10^{-8} $ \\ \hline 
			
			$e$ & $-2.5 \cdot 10^{-16} $ & $ \mathbin{\color{gray} 265} $ & 
			$ 2.2  \cdot 10^{-3}$ & - & $\mathbin{\color{gray}  9 \cdot 10^6}$ & $ 6 \cdot 10^{-9} $ \\ 
			\hline \hline
		\end{tabular}
	\end{center}
\end{table}

\twocolumngrid

\begin{figure}[h]
	\begin{center}
		\includegraphics[scale=0.9]{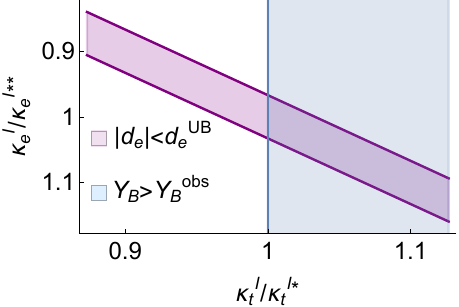}
	\end{center}
	\caption{\textit{The interplay between the electron and the top could allow the top to saturate the observed \ac{BAU} via $\kappa_t =1+ i \kappa_t^{I*}$, while canceling the contribution of $\kappa_t^{I*}$ to the \ac{eEDM} using $\kappa_e = 1+ i \kappa_e^{I**}$. Accordingly, $\kappa_e^{I**}$ is set such that $d_e$, i.e. the sum of equations~\eqref{BrodTopeEDM} and \eqref{eEDMAltmannshofer}, equals zero.	Since the next leading bound on both particles is orders of magnitude weaker, such cancellation would be difficult to detect.} }  \label{plot:etInterplay}
\end{figure}

%---------------------------
\paragraph*{Results} 
%---------------------------
We present our results in Table~\ref{tab:BAUresults}. The first prominent result is that no light charged fermion could give the dominant contribution to the \ac{BAU}. Of all the \ac{SM} charged fermions only the $\tau$ could produce $100 \%$ of the observed \ac{BAU}~\cite{deVries:2018tgs,Fuchs:2020uoc}. The next in importance can be the $\mu$~\cite{Fuchs:2019ore}, which brings us to consider the relatively negligible effect of the light quarks. In addition to their low contribution to the \ac{BAU}, the bounds on light quarks are comparable to these of the leptons, and thus their maximal percentage is relatively small.\\
That being said, one could consider two flavor modification, in which case the electron has a special feature. Assuming the numerical result is linear with multiple \ac{CPV} sources of different species, the electron could cancel the contribution of some particles to the \ac{eEDM}, while leaving the contribution to the \ac{BAU} essentially unchanged. When combined, the interference allows particles that are constrained mostly by the \ac{eEDM}, such as the top~\cite{Fuchs:2020uoc}, to account for the \ac{BAU}. For example, for $\kappa_t^{I*} \approx 0.04$, the top generates $Y_B^\text{obs}$, while $\kappa_e^{I**} \approx-0.06$ cancels the top's contribution to the \ac{eEDM} (See Fig.~\ref{plot:etInterplay}). Because of possible cancellation, it is much harder to exclude such an elusive hypothetical scenario.\\

%---------------------------
\paragraph*{Conclusions}
%---------------------------
We consider the \ac{CPV} source resulting from a dim-$6$ \ac{SMEFT} term which couples three Higgs fields to the \ac{SM} fermions. We apply the procedure described in Ref.~\cite{Fuchs:2020pun} to evaluate the complete set of single flavor modifications from all of the \ac{SM} charged fermions.\\
We deduce that although a larger \ac{dim-$4$} Yukawa coupling enhances the \ac{CPV} source, quarks have more washout than leptons and are therefore less favorable candidates to produce the \ac{BAU}. Moreover, to saturate $Y_B^\text{obs}$, some of the particles require non-perturbative \ac{dim-$4$} Yukawa couplings, e.g. the up and down quarks, and are therefore unequivocally ruled out as sole-contributors via this mechanism.\\
Constrained by \ac{U.B.}s of the electron and neutron \ac{EDM}s and measurements of various Higgs boson processes, we evaluate the maximal contribution from each single flavor modification (See Table~\ref{tab:BAUresults}). We conclude that the $\tau$ is the only candidate able to produce the observed \ac{BAU}~\cite{deVries:2018tgs,Fuchs:2020uoc}. Other than the $\mu$, which could provide up to $16 \%\  Y_B^\text{obs}$~\cite{Fuchs:2019ore}, the rest of the charged fermions produce negligible contributions (less than $ 6 \%\ Y_B^\text{obs}$).\\
An interplay of different flavors could relax current bounds to allow the observed baryon asymmetry be accounted for by a \ac{CPV} Yukawa coupling. Specifically, the interplay of the top and the electron could allow the top to saturate $Y_B^\text{obs}$. 
% In addition, off-diagonal couplings should also be considered in a more complete description.

\begin{acknowledgments}
	We are grateful to Yossi Nir and Yehonatan Viernik for helpful discussions.
\end{acknowledgments}

% \newpage
%---------------------------
\appendix
%---------------------------
\onecolumngrid

\begin{table}[t]
	\begin{center}
		\caption{ \it Collider limits on the signal strength $\mu_{h \to f \bar{f}}$ from which we evaluate the \ac{U.B.} at $\sim 95\%$~\ac{C.L.}. Combined with the \ac{SM} prediction for the branching ratio, $\mathcal{B}^\text{SM}$, we constrain $\left| \kappa_f \right|$ using Eqn.~\eqref{eqn:muFormula} for $\ell = \tau, \mu, e, b$, Eqn.~\eqref{eqn:muFormulab} for $q = u,c,d,s$ and Eqn.~\eqref{eqn:muFormulac} for the top. We extrapolated $\mathcal{B}^\text{SM}$ for the electron (up and down quarks) from that of the muon (strange), using $\mathcal{B}_{h \to f \bar{f}}^\text{SM} \propto m_f^2$.}
		\label{tab:colliderLimits}
		\renewcommand{\arraystretch}{1.5}
		\begin{tabular}{C{1.5cm} C{2cm} C{3.5cm} C{2cm} C{2.5cm}}  \hline\hline
			
			Channel & Experiment & $\mu_{h \to f \bar{f}}^\text{best fit}$& $\mu_{h \to f \bar{f}}^{\text{U.B.}}$& $\mathcal{B}_{h \to f \bar{f}}^\text{SM}$ \\ [2pt] \hline\hline
			
			% tau
			$h \to \tau \bar{\tau}$ &ATLAS+CMS& $0.91 \pm 0.13$ \cite{Fuchs:2020uoc} & $1.1$ & $6.3 \cdot 10^{-2}$ \cite{LHCHiggsCrossSectionWorkingGroup:2016ypw} \\ \hline
			
			% mu
			\multirow{2}{*}{$h \to \mu \bar{\mu}$} & ATLAS & $1.2 \pm 0.6$ \cite{Aad:2020xfq}& \multirow{2}{*}{$1.8$} & \multirow{2}{*}{$2.2 \cdot 10^{-4}$ \cite{LHCHiggsCrossSectionWorkingGroup:2016ypw} } \\ 
			& CMS & $1.19\pm 0.44$ \cite{ICHEP:2020}& \\ \hline
			
			% bottom
			$h \to b \bar{b}$ &CMS& $1.04 \pm 0.20$ \cite{Sirunyan:2018kst} & $1.4$ & $0.58$ \cite{LHCHiggsCrossSectionWorkingGroup:2016ypw} \\ \hline
			
			% top
			\multirow{2}{*}{$ggF + t\bar{t}h$} &\multirow{2}{*}{ ATLAS+CMS} & \multirow{2}{*}{$1.09 \pm 0.08$ \cite{Fuchs:2020uoc}} & \multirow{2}{*}{$1.2$} & $\mathcal{B} \leri{h \to gg}^\text{SM} = $ \\ 
			& & & & $8.2\cdot 10^{-2}$ \cite{LHCHiggsCrossSectionWorkingGroup:2016ypw} \\ \hline
			
			% charm
			$h \to c \bar{c}$ & \multicolumn{3}{c}{\multirow{4}{*}{using $\mu_{h \to b \bar{b}} \geq 0.71$ \cite{Sirunyan:2018kst}  }}& $2.9 \cdot 10^{-2}$ \cite{LHCHiggsCrossSectionWorkingGroup:2016ypw} \\ \cline{1-1} \cline{5-5}
			
			% strange
			$h \to s \bar{s}$ & & & & $4.40 \cdot 10^{-4}$ \cite{LHCHiggsCrossSectionWorkingGroup:2011wcg} \\  \cline{1-1} \cline{5-5}
			
			% down
			$h \to d \bar{d}$ &  & & & $1.1 \cdot 10^{-6}$ \cite{LHCHiggsCrossSectionWorkingGroup:2011wcg}
			% (from $\mathcal{B} \leri{h \to  s \bar{s}}^\text{SM}$) 
			\\  \cline{1-1} \cline{5-5}
			
			% up
			$h \to u \bar{u}$ &  & & & $ 2.4 \cdot 10^{-7}$ \cite{LHCHiggsCrossSectionWorkingGroup:2011wcg}
			% (from $\mathcal{B} \leri{h \to  s \bar{s}}^\text{SM}$) 
			\\ \hline
			
			% electron
			$h \to e \bar{e}$ &ATLAS & $\mathcal{B}^\text{U.B.} = 3.6 \cdot 10^{-4}$ \cite{ATLAS:2019old} &  $7.0 \cdot 10^{4} $ & $5.1 \cdot 10^{-9}$ \cite{LHCHiggsCrossSectionWorkingGroup:2016ypw} 
			% (from $\mathcal{B} \leri{h \to \mu \bar{\mu}}^\text{SM}$)  
			\\ \hline \hline
		\end{tabular}
	\end{center}
\end{table}

\newpage
\twocolumngrid

\section{Benchmark parameters} \label{benchParameters}
The input used for this work is the following:

\begin{itemize}
	\item Coupling constant at nucleation temperature~\cite{deVries:2017ncy}:
	\begin{align}
	g_s &= 1.23 \,, &
	g&= 0.65 \,, &
	g' &= 0.36 \,.
	\end{align}
	\item Bubble wall velocity and width~\cite{deVries:2018tgs}:
	\begin{align}
	v_w &= 0.05 \,, &
	L_w &= 0.11\ \text{GeV}^{-1}\,.
	\end{align}
	\item \ac{VEV} during nucleation~\cite{deVries:2017ncy} and at $0$ temperature:
	\begin{align}
	v_N &= 152 \text{ GeV}\,, &
	v_0 &= 246\text{ GeV}\,.	
	\end{align}
	\item The \ac{SM} fermion masses were taken from~\cite{Tanabashi:2018oca}. 
	\item The diffusion coefficients of leptons ($\ell$) and quarks ($q$) are~\cite{Joyce:1994zn}:
	\begin{align} \label{diffusionCoefficients}
	D_{\ell_L} &= D_{h} = \frac{100}{T}\,, & 
	D_{\ell_R} &= \frac{380}{T}\,,\nonumber \\
	D_{q_L} &= D_{q_R}= \frac{6}{T} \,.
	\end{align}
	\item The Mass and Yukawa rates are given in Table~\ref{tab:interactionRates}.
	\begin{table}[b]
	\caption{\textit{Relaxation rates (for the broken phase), and Yukawa rate (for both phases), calculated from~\cite{Fuchs:2020pun}.}}
	\label{tab:interactionRates}
	\begin{center}
		\renewcommand{\arraystretch}{2}
		\begin{tabular}{c| c| c}
			\hline 
			Particle & $\Gamma_M^{B}$ (GeV) & $\Gamma_Y$ (GeV) \\ \hline \hline
			$\tau$ & $4.9 \cdot 10^{-3} $ & $ 5.6 \cdot 10^{-4}$  \\ \hline
			$\mu$ & $1.7 \cdot 10^{-5} $ & $ 2.0\cdot 10^{-6}$  \\ \hline
			$e$ & $3.9 \cdot 10^{-10} $ & $ 4.4\cdot 10^{-11}$  \\ \hline
			$t$ & $102$ & $ 2.6 $  \\ \hline
			$c$ & $4.7 \cdot 10^{-3}$ & $1.6 \cdot 10^{-4}$  \\ \hline
			$u$ & $1.4 \cdot 10^{-8}$ & $4.7 \cdot 10^{-10}$  \\ \hline
			$b$ & $5.3 \cdot 10^{-2} $ & $ 1.7 \cdot 10^{-3}$  \\ \hline
			$s$ & $2.7 \cdot 10^{-5}$ & $9.0 \cdot 10^{-7}$  \\ \hline
			$d$ & $6.5 \cdot 10^{-8}$ & $2.1 \cdot 10^{-9}$  \\ \hline
		\end{tabular}
	\end{center}
\end{table}
	These interaction rates are calculated for $T_I = T_R = 0$.
	\item The weak sphaleron rate~\cite{Bodeker:1999gx}
	\begin{align}
	\Gamma_{ws} &= 6 \underbrace{\kappa}_{\sim 20} \underbrace{\alpha_w}_{\alpha_w = \frac{g^2}{4 \pi}}^5 T = 120 T \leri{ \frac{g^2}{4 \pi}} ^5 \nonumber \\
	& \overset{\text{nucleation}}{\longrightarrow} 120 T_N \leri{ \frac{0.65^2}{4 \pi}} ^5 \approx 4.5 \cdot 10^{-4}\ \text{GeV} \,.
	\end{align}
	\item The strong sphaleron rate~\cite{Moore:2010jd}
	\begin{align}
	\Gamma_{ss} &= 14 \alpha_s^4 T \approx 0.26\ \text{GeV}\,.
	\end{align}
	\item Thermal width~\cite{Elmfors:1998hh}:
	\begin{align}
	\Gamma_\text{leptons} &\approx 2 \cdot 10^{-3}\ T \,, &
	\Gamma_\text{quarks} &\approx0.16\ T \,.
	\end{align}
\end{itemize}
The temperature during nucleation is $T_N = 88$ GeV, and for the SM we use $\mathcal{R} = \frac{15}{4}$.\\

Thermal functions:
\begin{itemize}
	\item The source is given by~\cite{Lee:2004we,Cirigliano:2006wh}
	\begin{align} \label{sourceTerm}
	S_f (z;T) &= \frac{v_w N_c^f}{\pi^2} \Im \leri{m_f' m_f^*} J_f(T) \nonumber \\
	&=  \frac{v_w N_c^f {y^f_\text{SM}}^2 }{2 \pi^2 v_0^2} J_f(T)  h^3(z)  h'(z) \times \kappa_I^f  \,, \\
	J_f (T)&= \int\limits_0^\infty \frac{k^2 dk}{\omega_L^f \omega_R^f} \Im \left[ \frac{n_f \leri{\mathcal{E}_L^f} - n_f\leri{\mathcal{E}_R^{f *}}}{\leri{\mathcal{E}_L^f - \mathcal{E}_R^{f *}}^2 }  \leri{\mathcal{E}_L^f\mathcal{E}_R^{f *} -k^2}   \right. \nonumber \\
	& \left. + \frac{n_f \leri{\mathcal{E}_L^f} + n_f\leri{\mathcal{E}_R^{f}}}{\leri{\mathcal{E}_L^f + \mathcal{E}_R^{f}}^2 }  \leri{\mathcal{E}_L^f\mathcal{E}_R^{f} +k^2} \right] \,.
	\end{align}
	For the background Higgs boson field we use the kink solution:
	\begin{align} \label{kink sol}
	h &= \frac{v_N}{2} \leri{1 + \tanh(\frac{z}{L_w})} \,.
	\end{align}
	\item The frequencies, energies and Fermi- Dirac distributions are
	\begin{align}
	\omega_{L,R}^{f_i} (k) &= \sqrt{k^2 + \Re(\delta m_{f_{i\ L,R}}^2(T))}\,,  \nonumber \\
	\mathcal{E}_L^{f_i} &= \omega_{L,R}^{f_i}(k) -i \Gamma_{f_i} \,, \nonumber \\
	n_f \leri{\mathcal{E}} &= \frac{1}{e^{\frac{\mathcal{E}}{T}} +1 } % Fermi Dirac distribution
	\end{align}
	\item The thermal masses are given by~\cite{Enqvist:1997ff}
	\begin{align}
	\Re(\delta m_{L_{Li}}^2(T))&= \leri{\frac{3}{32} g^2 + \frac{1}{32} {g'}^{2} + \frac{1}{16} y_{e_i}^2} T^2 \equiv a_{L_{Li}}^2 T^2 \,, \nonumber \\
	\Re(\delta m_{e_{iR}}^2(T))&= \leri{ \frac{1}{8} {g'}^{2} + \frac{1}{8} y_{e_i}^2} T^2 \equiv a_{e_{Ri}}^2 T^2 \,, \nonumber \\
	\Re(\delta m_{Q_{Li}}^2(T))&= \left( \frac{1}{6} g_s^2 + \frac{3}{32} g^2 + \frac{1}{288} {g'}^{2} + \frac{1}{16} y_{u_i}^2 \right. \nonumber \\
	& \left. + \frac{1}{16} y_{d_i}^2 \right) T^2 \equiv a_{Q_{Li}}^2 T^2 \,, \nonumber \\
	\Re(\delta m_{u_{iR}}^2(T))&= \leri{ \frac{1}{6} g_s^2 +\frac{1}{18} {g'}^{2} + \frac{1}{8} y_{u_i}^2}  T^2 \equiv a_{u_{Ri}}^2 T^2\,, \nonumber \\
	\Re(\delta m_{d_{iR}}^2(T))&= \leri{ \frac{1}{6} g_s^2 +\frac{1}{72} {g'}^{2} + \frac{1}{8} y_{d_i}^2} T^2 \equiv a_{d_{Ri}}^2 T^2\,, \nonumber \\
	\Re(\delta m_{h}^2(T))&=  \left( \frac{3}{16} g^2 +\frac{1}{16} {g'}^{2} + \frac{1}{12} \sum_{i= e, \mu, \tau} y_{e_i}^2 \right. \nonumber \\
	& \left. + \frac{1}{4} \sum_{i= u,c,t} y_{u_i}^2 + \frac{1}{4} \sum_{i= d, s, b} y_{d_i}^2 \right) T^2 \equiv a_{h}^2 T^2 \,.
	\end{align}
	\item The finite temperature degrees of freedom are given by~\cite{deVries:2017ncy}
	\begin{align} \label{kFunction}
	k_{f_i} (a_{f_i}) &= k_0^{f_i} \frac{6}{\pi^2} \int \limits_{a_{f_i}}^\infty dx \frac{x e^x}{\leri{e^x \pm 1}^2} \sqrt{x^2-a_{f_i}^2}\,.
	\end{align}
	where $k_0^{f} $ is the number of degrees of freedom, and $+$ ($-$) is for fermions (the Higgs boson).
\end{itemize}

\section{Solving the set of transport equations: the two-step approach} \label{SolvingTransEqns}
In this appendix we summarize the analytic techniques used to calculate the produced baryon asymmetry, similarly to Ref.~\cite{Fuchs:2020pun}. We used the two-step approach: First we solve the set of transport equations, given in Eqn.~\eqref{transEqnFullSet}. Then, the \ac{BAU} is obtained by summing over the left-handed number densities and considering the weak sphaleron process. \\

\paragraph{Reduction of dimensions} The left hand side of Eqn.~\eqref{transEqnFullSet} can be written as a one dimensional - second order differential equation with respect to the bubble wall dimension denoted $z$~\cite{Cohen:1994ss}:
\begin{align} \label{diffApprox}
\partial f \equiv \partial_\mu f^\mu &=\frac{\partial f^0}{\partial t} - \vec{\nabla} \cdot \vec{f}
=\underbrace{\frac{\partial z}{\partial t}}_{\equiv v_w} \underbrace{\frac{\partial f^0}{\partial z}}_{f'} +  \vec{\nabla} \cdot \leri{ - D_f \vec{\nabla} f^0 } \nonumber \\
&= v_w f' - D_f \underbrace{\laplacian{f^0}}_{\equiv f''}
=v_w f' - D_f f'' \,,
\end{align}
where we used Fick's first law and the diffusion approximation.
\paragraph{Reduction of order}
We can solve this set of $N=16$- second order differential equations by reduction of order:
\begin{align}
g_{f_i} &\equiv f'_i \,, &
\vec{f} &= \begin{pmatrix}
\vec{U}_{R}& \vec{D}_{R} & \vec{Q}_{L}  & \vec{E}_{R} & \vec{L}_{L} &  h
\end{pmatrix} \,, \nonumber &
\vec{\chi} &= \begin{pmatrix}
\vec{f} & 
\vec{g}_f
\end{pmatrix} ^T \,.
\end{align}
\begin{align}
\vec{\chi}' = \begin{pmatrix} 0 &  \mathbbm{1} _{NxN} \\ \hat{\Gamma} & \hat{V} \end{pmatrix} \vec{\chi} +
\vec{S} \equiv \hat{K}  \vec{\chi} +
\vec{S} \,.
\end{align}
We are left with $2N$-first order differential equations. Then, $\hat{K} $ can be diagonalized numerically, e.g. by using MATLAB.

\paragraph{Numerical diagonalization}
The solution for the symmetric phase is given by
\begin{align}
{\vec{\chi}}^S &= \sum_{i=1}^{2N} C_i^S e^{\lambda_i^S z} \vec{u}_i^S  \equiv \hat{\Phi}^S (z) \vec{C}^S\,, \nonumber \\
\hat{\Phi}^X_{i,j} \leri{z} &= e^{\lambda_{jj}^X z} \leri{\vec{u}_j^X}_i \,.
\end{align}
where $C_j$'s are constants, $\lambda_j$ are the eigenvalues, and $\vec{u}_j$'s are the eigenvectors. We define
\begin{align}
\hat{\lambda} &= \text{diag}\leri{\lambda_i} & i&= [1 : 2N]\,. \\
\hat{\phi} &= \begin{pmatrix}
\vert & \vert & & \vert \\
\vec{u}_1 &\vec{u}_2 & \dots & \vec{u}_{2N} \\
\vert & \vert & & \vert
\end{pmatrix}
\end{align}
Accordingly,
\begin{align}
\hat{\Phi}_{i,j} \leri{z} &= \hat{\phi}_{ij} e^{\hat{\lambda}_{jj} z} 
= \begin{pmatrix}
\vert & \vert & & \vert \\
e^{\lambda_1 z} \vec{u}_1 & e^{\lambda_2 z} \vec{u}_2 & \dots & e^{\lambda_{2N} z} \vec{u}_{2N} \\
\vert & \vert & & \vert
\end{pmatrix}
\end{align}
The full solution in the broken phase is obtained by variation of parameters to be
\begin{align}
{\vec{\chi}}^B &= 
\hat{\Phi}^B (z) \vec{C}^B + \hat{\Phi}^B (z) \int\limits_0^z \leri{\hat{\Phi}^B (x) }^{-1} \vec{S} (x) \ dx \,.
\end{align}

\paragraph{Boundary conditions}
\begin{itemize}
	\item The integration constants of the divergent modes in the symmetric phase (correspond to $\lambda^S_j \leq 0$) are set to zero,
	\begin{align}
	C^{S}_{0-} &=0 \,.
	\end{align}
	\item The positive eigenvalues in the broken phase, $C^{B+}_j$ (correspond to $\lambda^B_j > 0$), are chosen such that they cancel the divergent part of the full solution at infinity:
	\begin{align}
	\vec{C}^{B}_+ &= -  \int\limits_0^\infty \leri{\hat{\Phi}^B_+ (x) }^{-1} \vec{S} (x) \ dx\,.
	\end{align}
	\item We demand continuity at $z=0$.
	\begin{itemize}
		\item In the symmetric phase we have
		\begin{align}
		\vec{\chi}^S_i \leri{z \to 0^-} &= \hat{\phi}^S_{ij} \vec{C}^S_j
		= \hat{\phi}^S_{i+} \vec{C}^S_+ \,.
		\end{align}
		\item In the broken phase we have
		\begin{align}
		\vec{\chi}^B_i \leri{z \to 0^+} &= \hat{\phi}^B_{ij} \vec{C}^B_j 
		=  \hat{\phi}^B_{i(0-)} \vec{C}^B_{0-} + \underbrace{ \hat{\phi}^B_{i+} \vec{C}^B_+ }_{\equiv b_i} \nonumber \\
		& = \hat{\phi}^B_{i(0-)} \vec{C}^B_{0-} + b_i \,. 
		\end{align}
	\end{itemize}
	Continuity is then
	\begin{align}
	\hat{\phi}^S_{i+} \vec{C}^S_+ &\overset{!}{=} \hat{\phi}^B_{i(0-)} \vec{C}^B_{0-} + b_i  \longrightarrow 
	\hat{\phi}^S_{i+} \vec{C}^S_+ - \hat{\phi}^B_{i(0-)} \vec{C}^B_{0-} = b_i \,.
	\end{align}
	We obtain a linear set of equations,
	\begin{align}
	\hat{\phi}_{SB} &\equiv \left(\begin{array}{c|c}
	\vert  & \vert \\
	\hat{\phi}^S_{i+}  & \hat{\phi}^B_{i(0-)}\\
	\vert & \vert
	\end{array} \right) \,, &
	\vec{C}_{SB} & \equiv \begin{pmatrix}
	\vec{C}^S_+  \\ - \vec{C}^B_{0-}
	\end{pmatrix} \,,  \nonumber \\
	& \hat{\phi}_{SB} \ \vec{C}_{SB} \overset{!}{=} b_i \,.
	\end{align}
	Solving it sets the rest of the coefficients.
\end{itemize}

\paragraph{The solution} The \ac{BAU} is then given by
\begin{align} \label{eqn:sphaleronProcess}
Y_B &= \frac{3 \Gamma_\text{ws}}{2 D_q \alpha_+ s} \int \limits_{0}^{-\infty} e^{-\alpha_- x} n_L(x)\ dx \,,
\end{align}
where $n_L$ is the density of left handed particles in the symmetric phase (where the weak sphaleron process is efficient), 
\begin{align}
n_L(z) &= \sum_{i=1}^3 \leri{ Q_{Li}(z) + L_{Li} (z) } \,,
\end{align}
and
\begin{align}
\alpha_\pm &= \frac{1}{2D_q} \leri{v_w \pm \sqrt{4 D_q \Gamma_\text{ws} \mathcal{R} + v_w^2}}\,.
\end{align}


\begin{thebibliography}{99}
	
 %\cite{Tanabashi:2018oca}
\bibitem{Tanabashi:2018oca}
M.~Tanabashi \textit{et al.} [Particle Data Group],
%``Review of Particle Physics,''
Phys. Rev. D \textbf{98}, no.3, 030001 (2018)
%doi:10.1103/PhysRevD.98.030001
%4856 citations counted in INSPIRE as of 15 Jun 2020	
 
 %\cite{Planck:2015fie}
 \bibitem{Planck:2015fie}
 P.~A.~R.~Ade \textit{et al.} [Planck],
 %``Planck 2015 results. XIII. Cosmological parameters,''
 Astron. Astrophys. \textbf{594}, A13 (2016)
 % doi:10.1051/0004-6361/201525830
 [arXiv:1502.01589 [astro-ph.CO]].
 %9991 citations counted in INSPIRE as of 25 Jul 2021

%\cite{Cline:2006ts}
\bibitem{Cline:2006ts}
J.~M.~Cline,
%``Baryogenesis,''
[arXiv:hep-ph/0609145 [hep-ph]].
%318 citations counted in INSPIRE as of 07 Feb 2022

%\cite{Morrissey:2012db}
\bibitem{Morrissey:2012db}
D.~E.~Morrissey and M.~J.~Ramsey-Musolf,
%``Electroweak baryogenesis,''
New J. Phys. \textbf{14}, 125003 (2012)
% doi:10.1088/1367-2630/14/12/125003
[arXiv:1206.2942 [hep-ph]].
%586 citations counted in INSPIRE as of 07 Feb 2022 

%\cite{Sakharov:1967dj}
\bibitem{Sakharov:1967dj} 
  A.~D.~Sakharov,
  %``Violation of CP Invariance, C asymmetry, and baryon asymmetry of the universe,''
  Pisma Zh.\ Eksp.\ Teor.\ Fiz.\  {\bf 5}, 32 (1967)
  [JETP Lett.\  {\bf 5}, 24 (1967)]
  [Sov.\ Phys.\ Usp.\  {\bf 34}, no. 5, 392 (1991)]
  [Usp.\ Fiz.\ Nauk {\bf 161}, no. 5, 61 (1991)].
  %doi:10.1070/PU1991v034n05ABEH002497
  %%CITATION = doi:10.1070/PU1991v034n05ABEH002497;%%
  %3136 citations counted in INSPIRE as of 27 Aug 2019
  
  %\cite{Gavela:1993ts}
  \bibitem{Gavela:1993ts}
  M.~B.~Gavela, P.~Hernandez, J.~Orloff and O.~Pene,
  %``Standard model CP violation and baryon asymmetry,''
  Mod. Phys. Lett. A \textbf{9}, 795-810 (1994)
  % doi:10.1142/S0217732394000629
  [arXiv:hep-ph/9312215 [hep-ph]].
  %411 citations counted in INSPIRE as of 07 Feb 2022
  
  %\cite{Huet:1994jb}
  \bibitem{Huet:1994jb}
  P.~Huet and E.~Sather,
  %``Electroweak baryogenesis and standard model CP violation,''
  Phys. Rev. D \textbf{51}, 379-394 (1995)
  % doi:10.1103/PhysRevD.51.379
  [arXiv:hep-ph/9404302 [hep-ph]].
  %495 citations counted in INSPIRE as of 07 Feb 2022
  
  %\cite{CMS:2019ybf}
  \bibitem{CMS:2019ybf}
  A.~M.~Sirunyan \textit{et al.} [CMS],
  %``Searches for physics beyond the standard model with the $M_\mathrm{T2}$ variable in hadronic final states with and without disappearing tracks in proton-proton collisions at $\sqrt{s}=$ 13 TeV,''
  Eur. Phys. J. C \textbf{80}, no.1, 3 (2020)
  % doi:10.1140/epjc/s10052-019-7493-x
  [arXiv:1909.03460 [hep-ex]].
  %79 citations counted in INSPIRE as of 07 Feb 2022
  
  %\cite{ATLAS:2019fgd}
  \bibitem{ATLAS:2019fgd}
  G.~Aad \textit{et al.} [ATLAS],
  %``Search for new resonances in mass distributions of jet pairs using 139 fb$^{-1}$ of $pp$ collisions at $\sqrt{s}=13$ TeV with the ATLAS detector,''
  JHEP \textbf{03}, 145 (2020)
  % doi:10.1007/JHEP03(2020)145
  [arXiv:1910.08447 [hep-ex]].
  %100 citations counted in INSPIRE as of 07 Feb 2022
  
%\cite{deVries:2017ncy}
\bibitem{deVries:2017ncy}
J.~de Vries, M.~Postma, J.~van de Vis and G.~White,
%``Electroweak Baryogenesis and the Standard Model Effective Field Theory,''
JHEP \textbf{01}, 089 (2018)
% doi:10.1007/JHEP01(2018)089
[arXiv:1710.04061 [hep-ph]].
%50 citations counted in INSPIRE as of 13 Jun 2021

%\cite{deVries:2018tgs}
\bibitem{deVries:2018tgs}
J.~De Vries, M.~Postma and J.~van de Vis,
%``The role of leptons in electroweak baryogenesis,''
JHEP \textbf{04}, 024 (2019)
%doi:10.1007/JHEP04(2019)024
[arXiv:1811.11104 [hep-ph]].
%11 citations counted in INSPIRE as of 26 May 2020

%\cite{Fuchs:2020uoc}
\bibitem{Fuchs:2020uoc}
E.~Fuchs, M.~Losada, Y.~Nir and Y.~Viernik,
%``$CP$ violation from $\tau$, $t$ and $b$ dimension-6 Yukawa couplings - interplay of baryogenesis, EDM and Higgs physics,''
JHEP \textbf{05}, 056 (2020)
%doi:10.1007/JHEP05(2020)056
[arXiv:2003.00099 [hep-ph]].
%3 citations counted in INSPIRE as of 29 Jun 2020

%\cite{Fuchs:2019ore}
\bibitem{Fuchs:2019ore}
E.~Fuchs, M.~Losada, Y.~Nir and Y.~Viernik,
%``Implications of the Upper Bound on $\boldsymbol{h\to\mu^+\mu^-}$ on the Baryon Asymmetry of the Universe,''
Phys. Rev. Lett. \textbf{124}, no.18, 181801 (2020)
%doi:10.1103/PhysRevLett.124.181801
[arXiv:1911.08495 [hep-ph]].
%2 citations counted in INSPIRE as of 19 Jun 2020

%\cite{Lee:2004we}
\bibitem{Lee:2004we}
C.~Lee, V.~Cirigliano and M.~J.~Ramsey-Musolf,
%``Resonant relaxation in electroweak baryogenesis,''
Phys. Rev. D \textbf{71}, 075010 (2005)
% doi:10.1103/PhysRevD.71.075010
[arXiv:hep-ph/0412354 [hep-ph]].
%136 citations counted in INSPIRE as of 13 Sep 2021

%\cite{Kuzmin:1985mm}
\bibitem{Kuzmin:1985mm}
V.~A.~Kuzmin, V.~A.~Rubakov and M.~E.~Shaposhnikov,
%``On the Anomalous Electroweak Baryon Number Nonconservation in the Early Universe,''
Phys. Lett. B \textbf{155}, 36 (1985)
%doi:10.1016/0370-2693(85)91028-7
%2990 citations counted in INSPIRE as of 11 Aug 2021

%\cite{Joyce:1994zn}
\bibitem{Joyce:1994zn}
M.~Joyce, T.~Prokopec and N.~Turok,
%``Nonlocal electroweak baryogenesis. Part 1: Thin wall regime,''
Phys. Rev. D \textbf{53}, 2930-2957 (1996)
% doi:10.1103/PhysRevD.53.2930
[arXiv:hep-ph/9410281 [hep-ph]].
%198 citations counted in INSPIRE as of 07 Feb 2022

%\cite{Fuchs:2020pun}
\bibitem{Fuchs:2020pun}
E.~Fuchs, M.~Losada, Y.~Nir and Y.~Viernik,
%``Analytic techniques for solving the transport equations in electroweak baryogenesis,''
JHEP \textbf{07}, 060 (2021)
% doi:10.1007/JHEP07(2021)060
[arXiv:2007.06940 [hep-ph]].
%4 citations counted in INSPIRE as of 13 Sep 2021

 %\cite{Cohen:1994ss}
\bibitem{Cohen:1994ss}
A.~G.~Cohen, D.~B.~Kaplan and A.~E.~Nelson,
%``Diffusion enhances spontaneous electroweak baryogenesis,''
Phys. Lett. B \textbf{336}, 41-47 (1994)
% doi:10.1016/0370-2693(94)00935-X
[arXiv:hep-ph/9406345 [hep-ph]].
%124 citations counted in INSPIRE as of 07 Feb 2022

%\cite{Guo:2016ixx}
\bibitem{Guo:2016ixx}
H.~K.~Guo, Y.~Y.~Li, T.~Liu, M.~Ramsey-Musolf and J.~Shu,
%``Lepton-Flavored Electroweak Baryogenesis,''
Phys. Rev. D \textbf{96}, no.11, 115034 (2017)
% doi:10.1103/PhysRevD.96.115034
[arXiv:1609.09849 [hep-ph]].
%41 citations counted in INSPIRE as of 11 Feb 2022

%\cite{Giudice:1993bb}
\bibitem{Giudice:1993bb}
G.~F.~Giudice and M.~E.~Shaposhnikov,
%``Strong sphalerons and electroweak baryogenesis,''
Phys. Lett. B \textbf{326}, 118-124 (1994)
% doi:10.1016/0370-2693(94)91202-5
[arXiv:hep-ph/9311367 [hep-ph]].
%94 citations counted in INSPIRE as of 13 Feb 2022

%\cite{Chung:2009cb}
\bibitem{Chung:2009cb}
D.~J.~H.~Chung, B.~Garbrecht, M.~J.~Ramsey-Musolf and S.~Tulin,
%``Lepton-mediated electroweak baryogenesis,''
Phys. Rev. D \textbf{81}, 063506 (2010)
% doi:10.1103/PhysRevD.81.063506
[arXiv:0905.4509 [hep-ph]].
%44 citations counted in INSPIRE as of 11 Feb 2022

 %\cite{Chien:2015xha}
\bibitem{Chien:2015xha}
Y.~T.~Chien, V.~Cirigliano, W.~Dekens, J.~de Vries and E.~Mereghetti,
%``Direct and indirect constraints on CP-violating Higgs-quark and Higgs-gluon interactions,''
JHEP \textbf{02}, 011 (2016)
% doi:10.1007/JHEP02(2016)011
[arXiv:1510.00725 [hep-ph]].
%79 citations counted in INSPIRE as of 14 Jun 2021

%\cite{Andreev:2018ayy}
\bibitem{Andreev:2018ayy}
V.~Andreev \textit{et al.} [ACME],
%``Improved limit on the electric dipole moment of the electron,''
Nature \textbf{562}, no.7727, 355-360 (2018)
%doi:10.1038/s41586-018-0599-8
%204 citations counted in INSPIRE as of 29 Jun 2020

%\cite{Panico:2018hal}
\bibitem{Panico:2018hal}
G.~Panico, A.~Pomarol and M.~Riembau,
%``EFT approach to the electron Electric Dipole Moment at the two-loop level,''
JHEP \textbf{04}, 090 (2019)
%doi:10.1007/JHEP04(2019)090
[arXiv:1810.09413 [hep-ph]].
%28 citations counted in INSPIRE as of 29 Jun 2020

%\cite{Brod:2018pli}
\bibitem{Brod:2018pli}
J.~Brod and E.~Stamou,
%``Electric dipole moment constraints on CP-violating heavy-quark Yukawas at next-to-leading order,''
[arXiv:1810.12303 [hep-ph]].
%16 citations counted in INSPIRE as of 14 Oct 2020

%\cite{Brod:2013cka}
\bibitem{Brod:2013cka}
J.~Brod, U.~Haisch and J.~Zupan,
%``Constraints on CP-violating Higgs couplings to the third generation,''
JHEP \textbf{11}, 180 (2013)
% doi:10.1007/JHEP11(2013)180
[arXiv:1310.1385 [hep-ph]].
%165 citations counted in INSPIRE as of 22 Dec 2020

%\cite{Altmannshofer:2015qra}
\bibitem{Altmannshofer:2015qra}
W.~Altmannshofer, J.~Brod and M.~Schmaltz,
%``Experimental constraints on the coupling of the Higgs boson to electrons,''
JHEP \textbf{05}, 125 (2015)
%doi:10.1007/JHEP05(2015)125
[arXiv:1503.04830 [hep-ph]].
%56 citations counted in INSPIRE as of 19 Jun 2020

%\cite{Abel:2020gbr}
\bibitem{Abel:2020gbr}
C.~Abel \textit{et al.} [nEDM],
%``Measurement of the permanent electric dipole moment of the neutron,''
Phys. Rev. Lett. \textbf{124}, no.8, 081803 (2020)
% doi:10.1103/PhysRevLett.124.081803
[arXiv:2001.11966 [hep-ex]].
%52 citations counted in INSPIRE as of 14 Oct 2020

%\cite{Brod:2018lbf}
\bibitem{Brod:2018lbf}
J.~Brod and D.~Skodras,
%``Electric dipole moment constraints on CP-violating light-quark Yukawas,''
JHEP \textbf{01}, 233 (2019)
% doi:10.1007/JHEP01(2019)233
[arXiv:1811.05480 [hep-ph]].
%15 citations counted in INSPIRE as of 14 Oct 2020

%\cite{LHCHiggsCrossSectionWorkingGroup:2016ypw}
\bibitem{LHCHiggsCrossSectionWorkingGroup:2016ypw}
D.~de Florian \textit{et al.} [LHC Higgs Cross Section Working Group],
%``Handbook of LHC Higgs Cross Sections: 4. Deciphering the Nature of the Higgs Sector,''
%doi:10.23731/CYRM-2017-002
[arXiv:1610.07922 [hep-ph]].
%1382 citations counted in INSPIRE as of 26 Aug 2021

%\cite{Sirunyan:2018kst}
\bibitem{Sirunyan:2018kst}
A.~M.~Sirunyan \textit{et al.} [CMS],
%``Observation of Higgs boson decay to bottom quarks,''
Phys. Rev. Lett. \textbf{121}, no.12, 121801 (2018)
%doi:10.1103/PhysRevLett.121.121801
[arXiv:1808.08242 [hep-ex]].
%200 citations counted in INSPIRE as of 21 Aug 2020

%\cite{Aad:2020xfq}
\bibitem{Aad:2020xfq}
G.~Aad \textit{et al.} [ATLAS],
%``A search for the dimuon decay of the Standard Model Higgs boson with the ATLAS detector,''
[arXiv:2007.07830 [hep-ex]].
%4 citations counted in INSPIRE as of 21 Aug 2020

%\cite{ICHEP:2020}
\bibitem{ICHEP:2020}
% Family name, INITIAL(S) (of the presenter). Year. Title of the presentation. Title of conference, date of conference, location of conference.
R.~Carlin [CMS],
% ``CMS highlights,''
40th International Conference on High Energy Physics (ICHEP2020),
3 Aug 2020,
virtual conference.
%``Proceedings, 40th International Conference on High Energy Physics (ICHEP2020),''

%\cite{LHCHiggsCrossSectionWorkingGroup:2011wcg}
\bibitem{LHCHiggsCrossSectionWorkingGroup:2011wcg}
S.~Dittmaier \textit{et al.} [LHC Higgs Cross Section Working Group],
%``Handbook of LHC Higgs Cross Sections: 1. Inclusive Observables,''
% doi:10.5170/CERN-2011-002
[arXiv:1101.0593 [hep-ph]].
%1633 citations counted in INSPIRE as of 29 Aug 2021

%\cite{ATLAS:2019old}
\bibitem{ATLAS:2019old}
G.~Aad \textit{et al.} [ATLAS],
%``Search for the Higgs boson decays $H \to ee$ and $H \to e\mu$ in $pp$ collisions at $\sqrt{s} = 13$ TeV with the ATLAS detector,''
Phys. Lett. B \textbf{801}, 135148 (2020)
% doi:10.1016/j.physletb.2019.135148
[arXiv:1909.10235 [hep-ex]].
%25 citations counted in INSPIRE as of 29 Aug 2021
 
 %\cite{Bodeker:1999gx}
 \bibitem{Bodeker:1999gx}
 D.~Bodeker, G.~D.~Moore and K.~Rummukainen,
 %``Chern-Simons number diffusion and hard thermal loops on the lattice,''
 Phys. Rev. D \textbf{61}, 056003 (2000)
 % doi:10.1103/PhysRevD.61.056003
 [arXiv:hep-ph/9907545 [hep-ph]].
 %212 citations counted in INSPIRE as of 13 Sep 2021
 
 %\cite{Moore:2010jd}
 \bibitem{Moore:2010jd}
 G.~D.~Moore and M.~Tassler,
 %``The Sphaleron Rate in SU(N) Gauge Theory,''
 JHEP \textbf{02}, 105 (2011)
 % doi:10.1007/JHEP02(2011)105
 [arXiv:1011.1167 [hep-ph]].
 %116 citations counted in INSPIRE as of 13 Sep 2021
 
 %\cite{Elmfors:1998hh}
 \bibitem{Elmfors:1998hh}
 P.~Elmfors, K.~Enqvist, A.~Riotto and I.~Vilja,
 %``Damping rates in the MSSM and electroweak baryogenesis,''
 Phys. Lett. B \textbf{452}, 279-286 (1999)
 %doi:10.1016/S0370-2693(99)00169-0
 [arXiv:hep-ph/9809529 [hep-ph]].
 %22 citations counted in INSPIRE as of 13 Sep 2021
 
 %\cite{Cirigliano:2006wh}
 \bibitem{Cirigliano:2006wh}
 V.~Cirigliano, M.~J.~Ramsey-Musolf, S.~Tulin and C.~Lee,
 %``Yukawa and tri-scalar processes in electroweak baryogenesis,''
 Phys. Rev. D \textbf{73}, 115009 (2006)
 % doi:10.1103/PhysRevD.73.115009
 [arXiv:hep-ph/0603058 [hep-ph]].
 %66 citations counted in INSPIRE as of 13 Sep 2021
 
 %\cite{Enqvist:1997ff}
 \bibitem{Enqvist:1997ff}
 K.~Enqvist, A.~Riotto and I.~Vilja,
 %``Baryogenesis and the thermalization rate of stop,''
 Phys. Lett. B \textbf{438}, 273-280 (1998)
 %doi:10.1016/S0370-2693(98)00963-0
 [arXiv:hep-ph/9710373 [hep-ph]].
 %24 citations counted in INSPIRE as of 13 Sep 2021
 

 
\end{thebibliography}
\end{document}